\documentclass[12pt,a4paper]{article}
\usepackage{rotating}
\usepackage{epsfig}

\begin{document}
\textwidth=135mm
 \textheight=200mm
\begin{center}
{\bfseries $Q^2$-evolution of parton densities at small $x$
values.\\
Charm contribution in the combined H1$\&$ZEUS $F_2$ data.
\footnote{{\small Talk at the International Session-Conference SNP PSD RAS
``Physics of Fundamental Interactions'', JINR, Dubna, April 12 - 15,
2016.}}}
\vskip 5mm
A. V. Kotikov 
 and B.G. Shaikhatdenov 
\vskip 5mm
{\small {\it 
Joint Institute for
Nuclear Research, 141980 Dubna, Russia}} \\
\end{center}
\vskip 5mm
\centerline{\bf Abstract}
The Bessel-inspired behavior of parton densities
at small $x$,
obtained in the case of the flat initial conditions for DGLAP evolution equations,
is used in the fixed flavor scheme
to analyze precise H1$\&$ZEUS combined data on the structure function $F_2$.

\vskip 10mm
A reasonable agreement between the HERA data  
and next-to-leading-order (NLO) approximation of
perturbative Quantum Chromodynamics (QCD) 
was observed for $Q^2 \geq 2$ GeV$^2$ 
\cite{CoDeRo},
whereby giving a glimmer of hope that
perturbative QCD is capable of describing the
evolution of parton densities
down to very low $Q^2$ values.

The ZEUS and H1 Collaborations have presented 
the new precise combined data \cite{Aaron:2009aa} on the 
structure function (SF) $F_2$. 
The aim of this short paper is to compare
the combined H1$\&$ZEUS data with 
the predictions obtained by using the so-called doubled asymptotic scaling (DAS)
approach~\cite{Q2evo}, 
which is in turn based on the analytical solutions to 
DGLAP equations
in the small-$x$ limit \cite{BF1}.

To improve the analysis at low $Q^2$ values, it is important to consider
well-known infrared modifications of the strong coupling 
constant. We are going to use the ``frozen'' and analytic versions (see,  
\cite{Cvetic1} and references therein).

The SF
$F_2$ is a sum of the light-quark $F_2^l$ and charm $F_2^c$ terms,
which look like  \cite{Q2evo,Illarionov:2008be}
\begin{eqnarray}
	F_2^l(x,Q^2) &=& e \, \bigl(f_q(x,Q^2) + \frac{2}{3}f a_s(Q^2)f_g(x,Q^2)\bigr), ~~
a_s(Q^2) = \frac{\alpha_s(Q^2)}{4\pi},  
\label{8a} \\
 F_2^c(x,Q^2) &=&
\frac{2}{3} \, e_c^2 a_s(Q^2+4m_c^2)
\, \left(1+ \frac{2(1-c)}{\sqrt{1+4c}} \ln \frac{\sqrt{1+4c}+1}{\sqrt{1+4c}-1}\right)
f_g(x,Q^2) ,
\nonumber
\end{eqnarray}
where
$e=(\sum_1^f e_i^2)/f$ is an average of light quark charges squared,  $c=m_c^2/Q^2$  and 
$m_c$ and $e_c=2/3$ are the mass and charge of a charm quark.
The SF $ F_2^c$ give contributions at $Q^2 \geq  m_c^2$.

The small-$x$ asymptotic expressions for parton densities $f_a$ 
can be written as follows
\footnote{Here, for simplicity we consider only  the 
leading order 
approximation
The NLO results can be found in  \cite{Q2evo}.}
\begin{eqnarray}
f_a(x,Q^2) &=& 
f_a^{+}(x,Q^2) + f_a^{-}(x,Q^2),~~(a=q,g) \nonumber \\
	f^{+}_g(x,Q^2) &=& \biggl(A_g + \frac{4}{9} A_q \biggl)
		I_0(\sigma) \; e^{-\overline d_{+} s} + O(\rho),~~
	f^{+}_q(x,Q^2) ~=~
\frac{f}{9} \frac{\rho I_1(\sigma)}{I_0(\sigma)}
+ O(\rho),
\nonumber \\
	f^{-}_g(x,Q^2) &=& -\frac{4}{9} A_q e^{- d_{-} s} \, + \, O(x),~~
	f^{-}_q(x,Q^2) 
~=~ A_q e^{-d_{-}(1) s} \, + \, O(x),
	\label{8.02}
\end{eqnarray}
where $I_{\nu}$ ($\nu=0,1$)  are the modified Bessel
functions,
\begin{equation}
s=\ln \left( \frac{a_s(Q^2_0)}{a_s(Q^2)} \right),~~
\sigma = 2\sqrt{\left|\hat{d}_+\right| s
 \ln \left( \frac{1}{x} \right)}  \ ,~~~ \rho=\frac{\sigma}{2\ln(1/x)} \ ,
\label{intro:1a}
\end{equation}
and 
\begin{equation}
\hat{d}_+ = - \frac{12}{\beta_0},~~~
\overline d_{+} = 1 + \frac{20f}{27\beta_0},~~~ 
d_{-} = \frac{16f}{27\beta_0}
\label{intro:1b}
\end{equation}
denote singular $\hat{d}_+$ and regular $\overline d_{+}$ parts of the anomalous dimensions 
$d_{+}(n)$ and $d_{-}(n)$, 
respectively, in the limit $n\to1$.

\begin{figure}[t]
\centering
\vskip 0.5cm
\includegraphics[height=0.75\textheight,width=1.05\hsize]{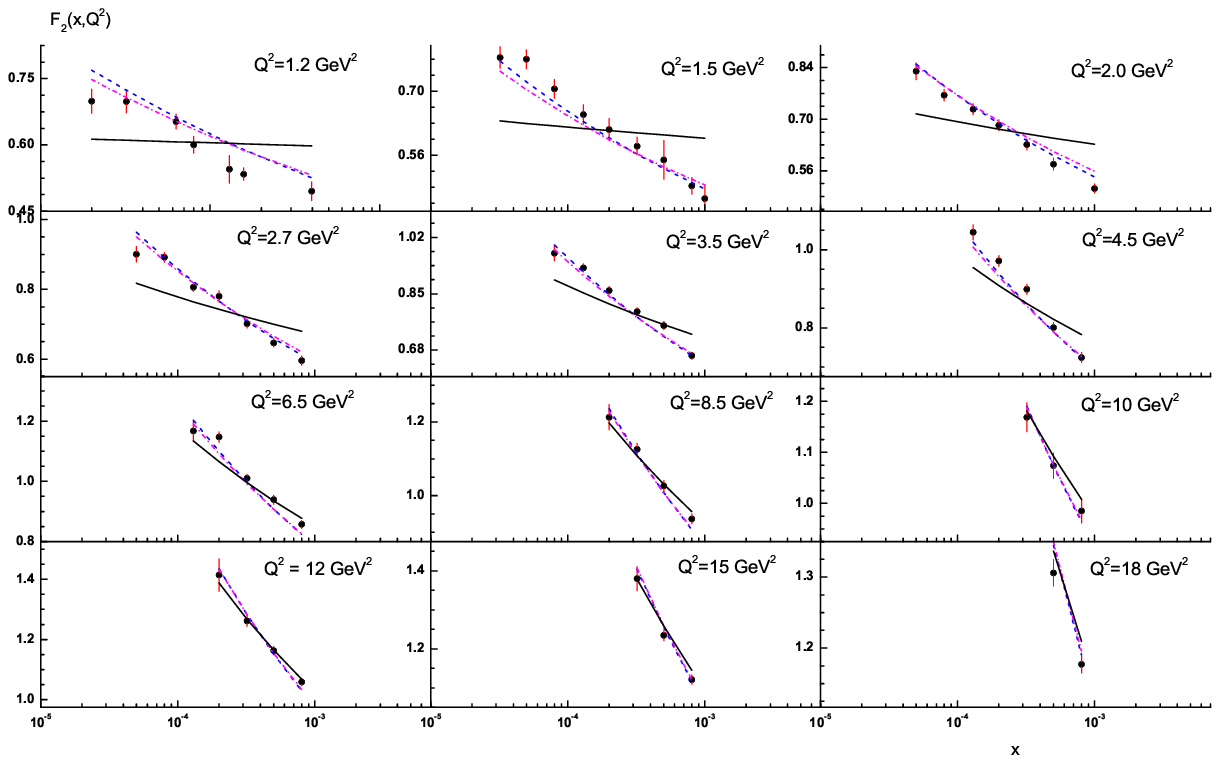}
\vskip -0.3cm
\caption{$x$ dependence of $F_2(x,Q^2)$ in bins of $Q^2$.
The combined experimental data from H1 and ZEUS Collaborations 
\cite{Aaron:2009aa} are
compared with the NLO fits for $Q^2\geq1$~GeV$^2$ implemented with the
standard (solid lines), frozen (dash-dotted lines), and analytic (dashed lines)
modifications of the strong coupling constant.}
\label{fig:F1}
\end{figure}

By using the above results 
we have
analyzed H1$\&$ZEUS data for $F_2$ \cite{Aaron:2009aa}.
Unlike our previous studies \cite{Kotikov:2012sm}, where we considered a charm quark to be light,
here we take the number of light quarks $f=3$, $m_c=1.275$ GeV and $\alpha_s(M^2_Z)=0.1168$ 
in agreement
with the
ZEUS results given in~\cite{H1ZEUS}.

As can be seen from Fig.~1, 
the twist-two approximation is reasonable for $Q^2 \geq 5$ GeV$^2$. 
At lower $Q^2$ we observe that the fits in the cases with ``frozen'' and
analytic strong coupling constants are very similar
(see also \cite{KoLiZo,Cvetic1,Kotikov:2012sm}) and describe fairly well the data in
the low $Q^2$ region, as opposed to the fit with a standard coupling constant.

As a next step in this direction, we are going to extend the dataset analyzed by
taking into consideration the combined H1$\&$ZEUS data for the SF
$F_2^c$ \cite{Abramowicz:1900rp} in the way adopted
in~\cite{Illarionov:2011km}, where old H1 and ZEUS data were used.\\

A.V.K. thanks the Organizing Committee of 
the International Session-Conference SNP PSD RAS
``Physics of Fundamental Interactions''
for invitation and support. This work was
supported in part by RFBR grant 16-02-00790-a.



\begin{thebibliography}{0}


%
\bibitem{CoDeRo} 
\textit{Cooper-Sarkar A. M., Abt I., Foster R., Wing M., Myronenko V. and 
Wichmann K.} //
  ``Study of HERA data at Low $Q^2$ and Low $x$,''
  arXiv:1605.08577 [hep-ph];
\textit{Kotikov A. V.} //
  Phys.\ Part.\ Nucl.\ 2007. {\bf V. 38.} P. 1
[Erratum-ibid.\  2007. {\bf V. 38.} P. 828].
%

\bibitem{Aaron:2009aa}
\textit{Aaron F. D. et al. (H1 and ZEUS Collaboration)} //
  JHEP. 2010. {\bf V. 1001.} P. 109.

%
\bibitem{Q2evo}\textit{Kotikov A. V. and Parente G} .//
Nucl. Phys. B. 1999. {\bf V. 549.} P. 242.;
%
%
\textit{Illarionov A. Yu.,
Kotikov A. V. and Parente G.} //
Phys. Part. Nucl. 2008. {\bf V. 39.} P. 307.



%
\bibitem{BF1}  %
\textit{De R\'ujula A.,
Glashow S. L., Politzer H. D., 
Treiman S. B., Wilczek F. and Zee A.} //  
Phys. Rev. D. 1974. {\bf V. 10.} P. 1649;
\textit{Ball R. D. and Forte S.} //
Phys. Lett. B. 1994. {\bf V. 336.} P. 77;
\textit{Mankiewicz L.,
Saalfeld A. and Weigl T.} //
Phys. Lett. B. 1997. {\bf V. 393.} P. 175.


\bibitem{Cvetic1}
\textit{Cvetic G.,
Illarionov A. Yu., Kniehl B. A. and Kotikov A. V.} //
 Phys. Lett. B. 2009. {\bf V. 679.} P. 350.

\bibitem{Illarionov:2008be}
\textit{Illarionov A. Yu., Kniehl B. A. and Kotikov A. V.} //
  Phys.\ Lett.\ B {\bf 663} (2008) 66


\bibitem{Kotikov:2012sm}
\textit{Kotikov A.B. and Shaikhatdenov B. G.} //
  Phys.\ Part.\ Nucl.\  {\bf 44} (2013) 543;
\textit{Kotikov A.B. and Shaikhatdenov B. G.} //
//  Phys.\ Atom.\ Nucl.\  {\bf 78} (2015) no.4,  525


%



%
\bibitem{H1ZEUS}  
\textit{Chekanov S. et al. (ZEUS Collaboration)} // Eur. Phys. J. C. 2001.
{\bf V. 21.} P. 443.





%
\bibitem{KoLiZo} \textit{ Kotikov A. V.,
Lipatov A. V. and Zotov N. P.} // 
 J.\ Exp.\ Theor.\ Phys.\ 2005.  {\bf V. 101.} P. 811;
\textit{Kotikov A. V., Krivokhizhin V. G. and Shaikhatdenov B. G.} //
  Phys.\ Atom.\ Nucl.\ 2012.  {\bf V. 75.} P. 507.

\bibitem{Abramowicz:1900rp}
 \textit{ Abramowicz H. et al. (H1 and ZEUS Collaboration)} //
  Eur.\ Phys.\ J.\ C {\bf 73} (2013) no.2,  2311


\bibitem{Illarionov:2011km}
\textit{Illarionov A. Yu. and Kotikov A. V.} //
  Phys.\ Atom.\ Nucl.\  {\bf 75} (2012) 1234;
 \textit{Kotikov A. V.} //
Proceedings Helmholtz International Summer School on
                        Physics of Heavy Quarks and Hadrons (HQ 2013),
pp. 184-194 
[Preprint DESY-PROC-2013-03].






\end{thebibliography}
\end{document}